\documentclass{article}
\usepackage[utf8]{inputenc}
\usepackage[english]{babel}
\pdfoutput=1
\usepackage{natbib}
\usepackage{graphicx}
\usepackage{amsmath}
\usepackage{named}
\usepackage{aas_macros}
\usepackage{authblk}
\begin{document}

\title{A Compact Full-disk Solar Magnetograph based on miniaturization of GONG instrument}

\author[1]{Sanjay Gosain}
\author[1]{Jack Harvey} 
\author[1]{Valentin Martinez-Pillet}
\author[2]{Tom Woods} 
\author[1]{Frank Hill}
\affil[1]{National Solar Observatory, Boulder, CO 80303, USA}
\affil[2]{Laboratory for Atmospheric and Space Physics, University of COlorado, Boulder, CO 80303, USA}

\date{Technical Report No.: NSO/NISP-2022-002}

\maketitle

\section{Abstract}
Designing compact instruments is the key for the scientific exploration by smaller spacecrafts such as cubesats or by deep space missions. Such missions require compact instrument designs to have minimal instrument mass. Here we present a proof-of-concept for miniaturization of the Global Oscillation Network Group (GONG) instrument. GONG instrument routinely obtains solar full-disk Doppler and magnetic field maps of the Sun's photosphere using Ni I 676 nm absorption line. A key concept for miniaturization of GONG optical design is to replace the bulky Lyot filter with a narrow-band interference filter and reduce the length of feed telescope. We present validation of the concept via numerical modeling as well as by  proof-of-concept observations.   

\section{Introduction}
Accurate mapping of solar surface magnetic and velocity fields is very important for: (a) characterizing the large-scale spatio-temporal patterns related to solar cycle and their interpretation in terms of the solar dynamo models \citep{Hathaway2015,charbonneau2010}, (b) force-free extrapolations to model global corona, \citep{Mackay2012,Wiegelmann2012}, (c) evolution of magnetic flux in solar active regions in relation to flares and CMEs \citep{vandriel2015,Shibata2011}. However, from the Earth-based observatories, or, from any single vantage point observation, we can only see one side of the Sun. Solar rotation allows us to eventually see the far-side, however, by that time the solar surface magnetic flux has evolved \citep{Sheeley2005}. A single vantage point observation also limits our space-weather modeling capabilities due to limited information about the surface boundary conditions.  This impacts the model predictions of the coronal and heliospheric magnetic field and the solar wind parameters \citep{Pevtsov2020}. Further, the obscure view of the solar polar regions from the ecliptic prevents us from mapping the high latitude magnetic fields, which are important for solar dynamo studies. 

Obtaining full disk measurements from different vantage points, accessible via space missions, such as from  Lagrange points (L4, L5) and/or polar orbits can provide the missing piece of information about the solar magnetic and velocity fields and help in understanding of the solar activity via better initialized models \citep{Gibson2018}. Solar Orbiter (SolO) mission \citep{Solanki2020}, which will observe the Sun from out of the ecliptic, is a recent example. 

Going to these new vantage points, however, can be very expensive and it is highly desirable to have compact and light-weight instruments to obtain the observations.  While there are other compact instrument designs such as magneto-optical filter (MOF) based concept \citep{Cacciani1978} or a tunable Fabry-Perot (FP) concept \citep{Berrilli2011}. However, their operation can add complexity, such as, strong magnetic fields in MOF can reduce the magnetic cleanliness of the spacecraft environment, an important criterion for complementary instruments such as in-situ magnetometer. While the accurate tuning of FP pass-band can be quite challenging when sampling narrow photoshpheric spectral lines, especially in the presence of a large spacecraft velocity relative to the Sun.

In this paper we describe present a concept for miniaturization of the GONG instrument \citep{Harvey1996, Harvey1998}. The GONG technique is simple to implement and interpretation of observable in terms of magnetic and velocity signals is straightforward. We discuss the miniaturization concept design and present proof-of-concept by using numerical modeling and obtaining sample observations at one of the GONG sites. 

\section{Brief description of GONG}
The operating principle of GONG is described in detail elsewhere \cite{Evans1981,Brown1981,Shepherd1993,Harvey1996}. In summary, GONG employs a wide-field polarizing cube Michelson interferometer \citep{Title1980} to perform phase-shifting interferometry and derives Doppler and Zeeman shift of the solar spectral line(s).

In practice, the path difference of the Michelson interferometer is changed from its nominal value in three steps, causing a phase-stepping of 2$\pi$/3 radians in the sinusoidal fringe pattern of the Michelson interferometer. When these fringes are superposed on a spectrally isolated (i.e., via a band-pass filter) portion of the solar spectra containing the target absorption lines, the resultant intensity is modulated. The phase of this modulation is proportional to the wavelength position of the solar absorption line. 

In case of GONG the phase-stepping is performed by means of a continuously rotating half waveplate (RHWP). The intensity on the detector is then modulated according to the angle of the RHWP, which is  given by the rotation rate $\omega$ of the RHWP and time, $t$, at which the intensity is sampled. 
\begin{equation}
I(t)=I_0 [1 + Mcos(4\omega t -\phi)]
\end{equation}

By sampling the modulated intensity signal at three Michelson phase-steps, $2\pi/3$ radians apart, one can determine, $I_0$, the mean intensity, $M$ the modulation amplitude, and $\phi$ the phase of the modulated signal. Of these, the $I_0$ and $M$ are related to the brightness and equivalent depth of the spectral line, respectively, while $\phi$ is proportional to the Doppler shift of the line.

The expressions for phase, $\phi$, and modulation amplitude, $M$, in terms of the three measured intensities, $I_1$, $I_2$ and $I_3$ can be written as:

\begin{equation}
tan~\phi = \sqrt{3}\frac{I_2 - I_3}{I_2 + I_3 - 2 I_1},
\end{equation}
and
\begin{equation}
M = \frac{1}{I_0} \sqrt{\frac{2}{3} \sum_{i=1}^{3} (I_i -I_0)^2} 
\end{equation}

where, $I_0$, is simply the average of the three intensities. 
The Doppler shift, $v$, of the spectral line is is related to the measured phase, $\phi$, as follows:

\begin{equation}
v = c \frac{\delta \phi}{\phi_0} 
\end{equation} 

where, $c$ is the speed of light and $\phi_0$ is the nominal phase difference corresponding to the optical path difference between the two arms of the  Michelson interferometer.   

To measure the line-of-sight (LOS) magnetic field, one simply measures the velocity of the left- and right- circularly polarized, Zeeman-split $\sigma-$ components of the spectral line, and takes the difference. The velocity difference is directly proportional to the LOS magnetic field, given by the expression:
\begin{equation}
B_{los}=\frac{0.5\Delta v}{4.67\times10^{-13}g_{eff}\lambda_0 c}
\end{equation}
where, $\Delta v$ is the Zeeman splitting in velocity units (in m/s), $g_{eff}$=1.42 is the effective Lande $g$-factor for Ni I line, $\lambda_0$=6767.8 is the rest wavelength (in \AA), and $c$ is the velocity of light (in m/s).
For these values we get the relation, $B_{los}$=0.37$\Delta v$.

\begin{figure}[ht]
\begin{center}
\includegraphics[scale=0.5]{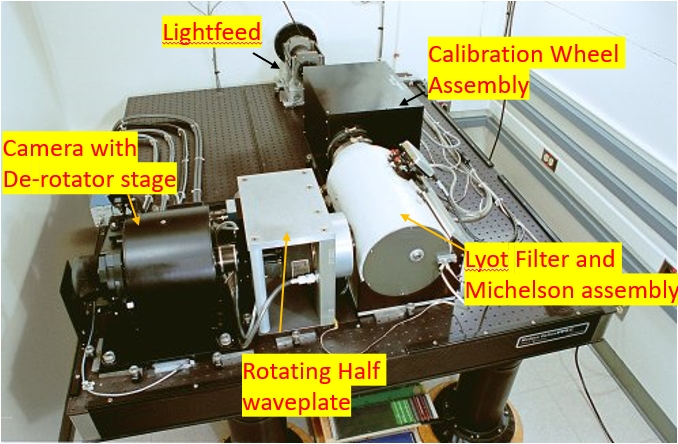}
\caption{The actual GONG optical bench setup with major components labeled is shown. The two mirror turret system that feeds the telescope tube is outside the housing and are not shown here. The size of the optical bench is 6 by 4 square feet. The light feed is a 1-m focal length telescope tube. The calibration wheel assembly contains the optics for flat field, dark, a liquid crystal retarder for polarimetry. The Lyot filter assembly is about a foot long train of birefringent elements, a 5 Angstroms wide interference filter, and polarizing Michelson cube. A rotating halfwaveplate is used to modulate the interferometer phase. The imaging camera is housed in a rotating stage to compensate for solar image rotation during the day. }
\label{fig:gongbench}
\end{center}
\end{figure}

To provide a perspective on the GONG instrument setup and its scale we show an image of the actual GONG setup in Figure~\ref{fig:gongbench}. 

\section{Miniaturization concept for GONG}
The key concept to make the current GONG design compact is to replace a rather bulky Lyot filter \citep{Lyot1944,Evans1949} with a narrow band interference filter. This replacement of Lyot, not only reduces 
the system length (and volume) but also saves a lot in terms of the mass of the system which is a crucial design parameter for deep space missions. However, replacing Lyot filters with interference filters comes with a small compromise. Lyot filters are known to have a large acceptance angle as compared to the inteference filters. This has two effects: (i) the pass-band of the interference filter shifts towards shorter wavelengths with increasing angle of incidence, and (ii) the system throughput of interference filter is smaller as compared to Lyot filter of same aperture size.

However, these two effects can be mitigated as follows. The pass-band shift with field angle is a smooth function which can be modeled quite accurately and calibrated using sunlight itself, i.e., by feeding the instrument a spatially (or disk)  averaged solar light, for example, by means of a small angle optical diffuser. Since, every pixel in the detector field-of-view sees same disk-averaged solar spectrum, the measured phase shift across field-of-view in the detector plane is then a measure of field dependence of narrow-band filter's pass-band shift. Such dependence can be measured accurately and modeled for post-facto correction in the measurements.  

The second effect of throughput limitation is more benign. GONG is not a photon starved instrument. This is  because the amount of solar radiation through rather large passband ($\sim$1\AA) of interference filter is dominated by the continuum radiation and is quite large than the typical detector pixel full-well can handle. Calculations show that typical exposure times of couple of milliseconds are enough to fill the full-well.

\subsection{Proof of concept: Numerical Modeling}
We made a simple numerical model of GONG instrument to verify the performance of commercially available interference filters. The instrument model was made as follows: Let I($\lambda$), P($\lambda$), T($\lambda$) be the solar absorption spectra, narrow-band prefilter and Michelson interferometer's transmission profile, respectively.

I($\lambda$) is approximated with the quiet Sun disk-center solar reference spectra from NSO Kitt Peak atlas. The wavelength range considered was $\pm$20\AA centered on the absorption line used in GONG, i.e., Ni I 6768\AA. 

\begin{figure}[ht!]
\begin{center}
\includegraphics[width=\textwidth]{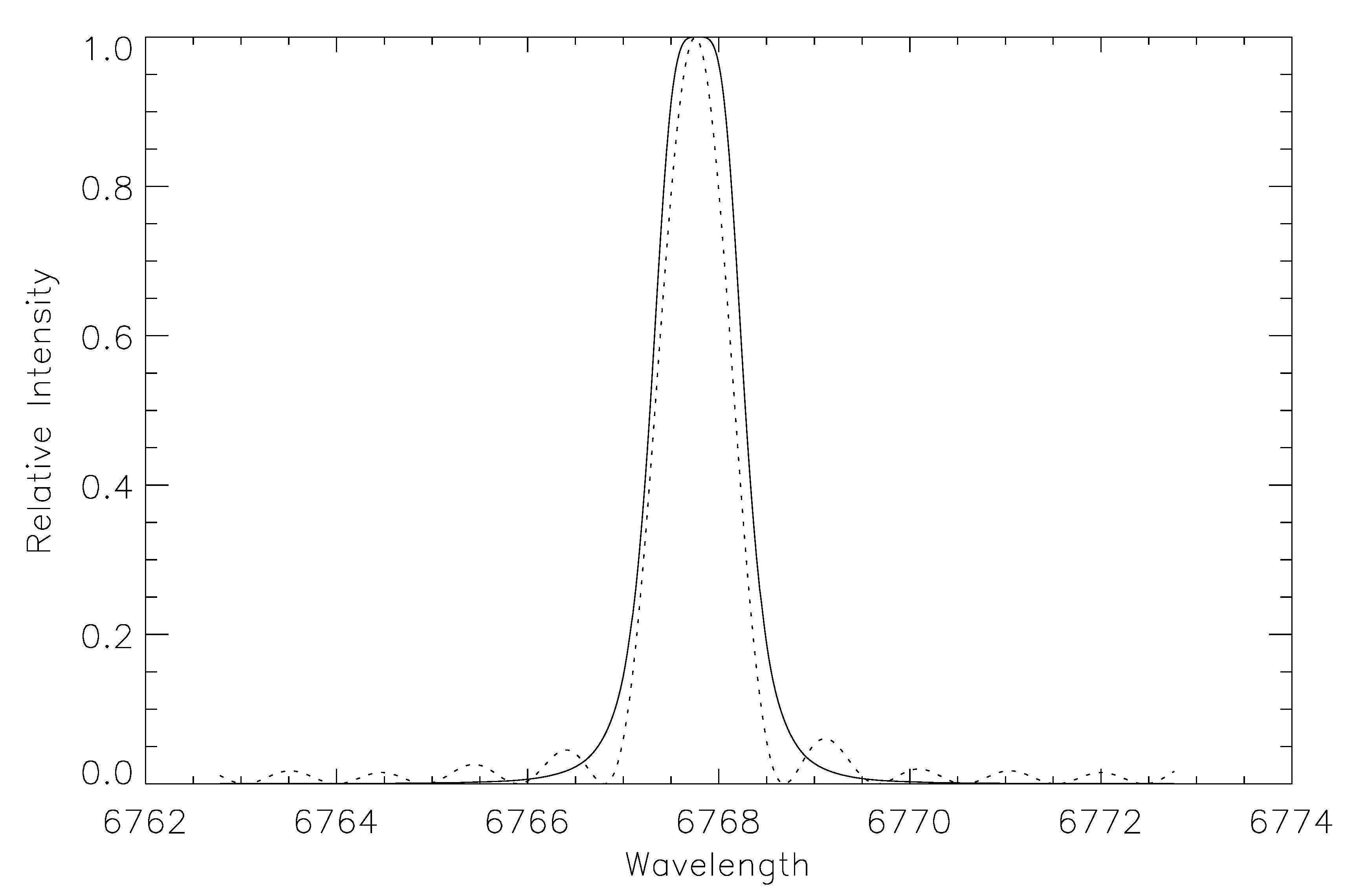}
\caption{The spectral profile, P($\lambda$), computed as per Eqn.(6) for the interference filter is shown with solid curve. The dotted curve shows the transmission profile computed for the current the GONG Lyot filter.}
\label{fig:isolator}
\end{center}
\end{figure}

The prefilter profile, P($\lambda$), was approximated using the following equation:
\begin{equation}
P(\lambda)=T_{max}\frac{1}{ 1+\left[\frac{2(\lambda-\lambda_0)}{\Delta\lambda} \right]^{2n}}  
\end{equation}
where, $\lambda_0$ is the central wavelength of the pass-band and $\Delta\lambda$ is the full-width at half maxima (FWHM) of the interference filter, which was set to 1\AA~here. Further, $n$ is the number of cavities in the interference filter, we have used $n$=1 here. Figure~\ref{fig:isolator} shows the transmission profile of modeled interference filter with two cavity design and pass-band of 1\AA (the solid curve). For reference we also show the computed transmission profile of the Lyot filter used by current GONG instrument (the dotted curve). 

\begin{figure}[ht!]
\begin{center}
\includegraphics[width=\textwidth]{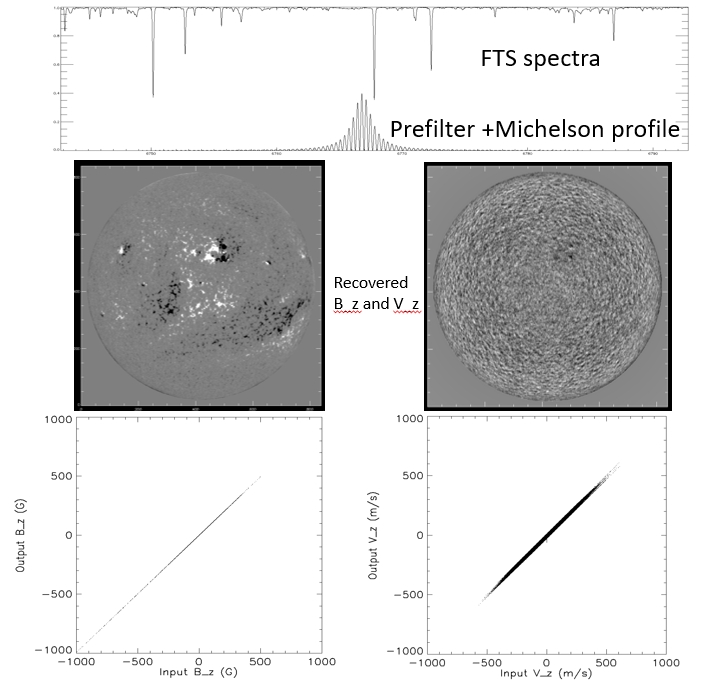}
\caption{Results of numerical simulations are shown here. Top panel shows spectral profiles of the solar absorption spectra (FTS atlas spectra) and instrumental transmission profile, i.e., a product of narrow-band interference filter and Michelson transmission profiles, at one of the three path difference setting of the Michelson interferometer. Middle panel shows the simulated measurements of the magnetic and velocity field of the full disk of the Sun. Bottom panel shows the scatter plots between the ``ground truth" GONG observations and simulated measurements.}
\label{fig:simula}
\end{center}
\end{figure}

Michelson transmission profile as a function of wavelength and phase delay, $d\phi$, is given as:
\begin{equation}
T(\lambda,d\phi)=0.5\left[1+cos(2\pi\Delta/\lambda +d\phi)\right]    
\end{equation}
where, $\Delta$ is the nominal optical path difference (OPD) of the Michelson interferometerfor the  Michelson used in GONG this is taken to be $\Delta$=1.5cm.

We can then simulate the intensity as measured by GONG, as follows: 
\begin{equation}
I_i=\Sigma_{i=1}^3 I(\lambda)P(\lambda)T(\lambda,d\phi_i)    
\end{equation}
where $d\phi_i$=i$2\pi/3$, for i=1,2 and 3, respectively. Using these three modeled intensities together with Eqns [2]-[5], we can derive the phase shift of the solar spectra. However, to create the magnetic maps of the Sun using Zeeman diagnostics we need two sets of these measurements, one each in left- and right- circularly polarized light. So, this makes a total of six observables.   

In reality, I($\lambda$) varies spatially across the solar disk due to changing physical parameters. Here we just focus on changing velocity and magnetic field across the Sun. We use full disk maps of line-of-sight velocity, $V_{los}$, and magnetic field, $B_{los}$, as observed by GONG, as the "ground truth". These ``ground truth" values, $B_{los}(x,y)$ and $V_{los}(x,y)$, are then used to produce synthetic spectra, $I_{\lambda}^{syn}(x,y)$. Using these input spectra, synthetic observables are produced using eqn.(8). Finally, from these synthetic observables we recover magnetic and velocity field using eqns.[2]-[5]. 

The comparison of input ``ground truth" with recovered values is made via scatter plots as shown in the bottom panels of Figure~\ref{fig:simula}. We find a good correlation between the input and output values of the Doppler and magnetic signals. These simulations suggest feasibility of replacing Lyot filter with a narrow-band interference filter of similar bandwidth. 

\begin{figure}[ht!]
\begin{center}
\includegraphics[width=\textwidth]{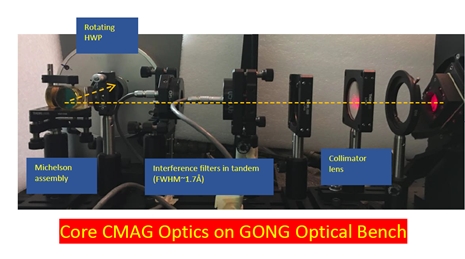}
\caption{The proof-of-concept demonstration setup at GONG bench is shown.}
\label{fig:poclabel}
\end{center}
\end{figure}

\section{Proof-of-concept: Observations}

\begin{figure}[ht!]
\begin{center}
\includegraphics[scale=0.7]{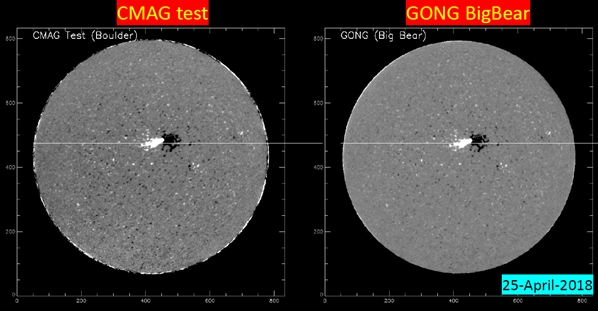}
\includegraphics[scale=0.99]{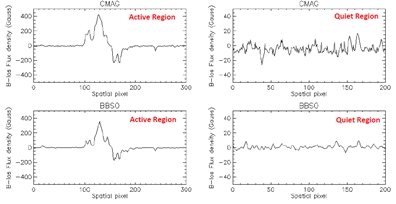}
\caption{Top row shows simultaneous magnetograms obtained from compact magnetograph (CMAG)  and GONG Big Bear site. Bottom panels show comparison of flux values along the white line marked over the fulldisk magnetograms, emphasizing active regions in left panels and quiet sun region in right panels.}
\label{fig:pocdata}
\end{center}
\end{figure}

\begin{figure}[ht!]
\begin{center}
\includegraphics[scale=0.9]{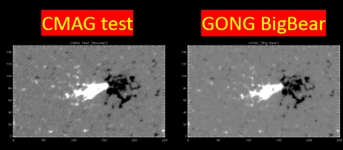}
\includegraphics[scale=0.95]{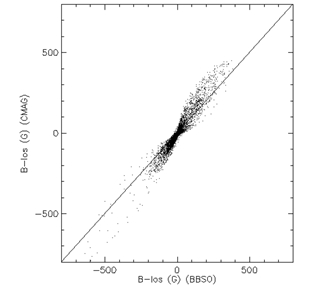}
\caption{ Top row shows active region magnetic map by CMAG and GONG. Bottom panels shows a scatter-plot of magnetic flux densities between the two measurements, pixel-by-pixel.  }
\label{fig:pocdataroi}
\end{center}
\end{figure}

For a quick demonstration of the proof-of-concept we replaced the Lyot filter with narrow-band interference filter in the GONG setup. The demonstration setup at GONG bench is shown in Figure~\ref{fig:poclabel}. A lens shown on the right collimates the primary solar image and through two single-cavity interference filters housed in temperature controlled cells, mounted in tandem followed by polarization analyzer and a Michelson interferometer cube. The beam passes through a quarter wave-plate on the exit side of the Michelson cube (not seen here) before entering a rotating half wave-plate assembly.

A full disk solar magnetogram was obtained on 25 April 2018 by using the measurements taken with the proof-of-concept setup shown in Figure~\ref{fig:poclabel}. We compare these observations with a simultaneous magnetogram obtained by GONG site at Big Bear, California, USA. Side-by-side comparison of the two magnetograms is shown in the top panel of Figure~\ref{fig:pocdata}. We compared the magnetic signals in the two magnetograms along the while line marked in top panel. We emphasize in the left panels the magnetic flux over the active region located near disk center, while similar comparison in the quiet region is shown in the right panels. 

A more detailed comparison of the magnetic signals is shown in Figure~\ref{fig:pocdataroi}, where we focus only on the strong active region fluxes. A scatter-plot in the bottom panel shows that the magnetic field values compare well given the  different atmospheric conditions in the two sites which lead to different flux distributions in pixels.  

These proof-of-concept observations and our numerical modeling demonstrates that using interference filters in place of Lyot filters is a viable alternative that leads to miniaturization of GONG instrument design. 

\section{Discussion and Conclusions}
We have demonstrated a miniaturization concept for GONG whereby we replace the bulky Lyot filter element with a narrow band interference filter. This allows one to design a very compact version of GONG optical design. Further, we have verified the proof-of-concept with numerical simulations and test observations to demonstrate proof-of-concept.

State-of-the-art interference filters with pass-band as narrow as $\sim$0.1 nm can be commercially obtained \citep{vmp2011}. The interference filters are very small as compared to the Lyot filter in size, only a few millimeters in thickness. Such filters have flown in space instruments and are typically  designed with hard coatings on radiation resistant substrates. 

On the other hand, the acceptance angle of bi-refringent filter is much higher than interference filter. For example, for a 1 degree deviation from normal angle of incidence the pass-band of field-widened \citep{Title1980} Lyot filter shifts only $\sim$20 m\AA, whereas for a high refractive index coating the corresponding shift for interference filter would be $\sim$230 m\AA. Thus, using interference filters would require slower optical beam and a good calibration of the wavelength response of the instrument with field-angle. However, the advantages of a narrowband interference filter over Lyot filter, namely, much smaller size (significant mass savings) far outweigh the acceptance angle disadvantage, which can be accurately calibrated and modeled in the laboratory as well as in-flight using solar light itself. 

Additional simplification in the GONG design is also possible by replacing the rotating half wave-plate (RHWP) with an electrically tunable liquid crystal variable retarder (LCVR). Using LCVR one can avoid the complication of using a mechanical system for precision rotation of the wave-plate, hence reducing the complexity and cost in the design. 

\bibliographystyle{plainnat}
\bibliography{references}

\begin{thebibliography}{21}
\providecommand{\natexlab}[1]{#1}
\providecommand{\url}[1]{\texttt{#1}}
\expandafter\ifx\csname urlstyle\endcsname\relax
  \providecommand{\doi}[1]{doi: #1}\else
  \providecommand{\doi}{doi: \begingroup \urlstyle{rm}\Url}\fi

\bibitem[{Berrilli} et~al.(2011){Berrilli}, {Cocciolo}, {Giovannelli}, {Del
  Moro}, {Giannattasio}, {Piazzesi}, {Stangalini}, {Egidi}, {Cavallini},
  {Greco}, and {Selci}]{Berrilli2011}
Francesco {Berrilli}, Martina {Cocciolo}, Luca {Giovannelli}, Dario {Del Moro},
  Fabio {Giannattasio}, Roberto {Piazzesi}, Marco {Stangalini}, Alberto
  {Egidi}, Fabio {Cavallini}, Vincenzo {Greco}, and Stefano {Selci}.
\newblock {The Fabry-Perot interferometer prototype for the ADAHELI solar small
  mission}.
\newblock In Silvano {Fineschi} and Judy {Fennelly}, editors, \emph{Solar
  Physics and Space Weather Instrumentation IV}, volume 8148 of \emph{Society
  of Photo-Optical Instrumentation Engineers (SPIE) Conference Series}, page
  814807, October 2011.
\newblock \doi{10.1117/12.893552}.

\bibitem[{Brown}(1981)]{Brown1981}
T.~{Brown}.
\newblock {The Fourier Tachometer: Principles of Operation and Current Status}.
\newblock In Richard~B. {Dunn}, editor, \emph{Solar instrumentation: What's
  next?}, page 150, March 1981.

\bibitem[{Cacciani} and {Fofi}(1978)]{Cacciani1978}
A.~{Cacciani} and M.~{Fofi}.
\newblock {The Magneto-Optical Filter. II. Velocity Field Measurements}.
\newblock \emph{\solphys}, 59\penalty0 (1):\penalty0 179--189, September 1978.
\newblock \doi{10.1007/BF00154941}.

\bibitem[{Charbonneau}(2010)]{charbonneau2010}
Paul {Charbonneau}.
\newblock {Dynamo Models of the Solar Cycle}.
\newblock \emph{Living Reviews in Solar Physics}, 7\penalty0 (1):\penalty0 3,
  September 2010.
\newblock \doi{10.12942/lrsp-2010-3}.

\bibitem[{Evans}(1981)]{Evans1981}
J.~{Evans}.
\newblock {The Fourier Tachometer. The Solid Polarizing Interferometer}.
\newblock In Richard~B. {Dunn}, editor, \emph{Solar instrumentation: What's
  next?}, page 155, March 1981.

\bibitem[{Evans}(1949)]{Evans1949}
John~W. {Evans}.
\newblock {The birefringent filter}.
\newblock \emph{Journal of the Optical Society of America (1917-1983)},
  39\penalty0 (3):\penalty0 229, March 1949.

\bibitem[{Gibson} et~al.(2018){Gibson}, {Vourlidas}, {Hassler}, {Rachmeler},
  {Thompson}, {Newmark}, {Velli}, {Title}, and {McIntosh}]{Gibson2018}
Sarah~E. {Gibson}, Angelos {Vourlidas}, Donald~M. {Hassler}, Laurel~A.
  {Rachmeler}, Michael~J. {Thompson}, Jeffrey {Newmark}, Marco {Velli}, Alan
  {Title}, and Scott~W. {McIntosh}.
\newblock {Solar Physics from Unconventional Viewpoints}.
\newblock \emph{Frontiers in Astronomy and Space Sciences}, 5:\penalty0 32,
  September 2018.
\newblock \doi{10.3389/fspas.2018.00032}.

\bibitem[{Harvey} and {GONG Team}(1998)]{Harvey1998}
J.~{Harvey} and {GONG Team}.
\newblock {GONG instrument and science}.
\newblock \emph{Bulletin of the Astronomical Society of India}, 26:\penalty0
  135, January 1998.

\bibitem[Harvey et~al.(1996)Harvey, Hill, Hubbard, Kennedy, Leibacher, Pintar,
  Gilman, Noyes, Title, Toomre, Ulrich, Bhatnagar, Kennewell, Marquette,
  Patr{\'o}n, Sa{\'a}, and Yasukawa]{Harvey1996}
J.~W. Harvey, F.~Hill, R.~P. Hubbard, J.~R. Kennedy, J.~W. Leibacher, J.~A.
  Pintar, P.~A. Gilman, R.~W. Noyes, A.~M. Title, J.~Toomre, R.~K. Ulrich,
  A.~Bhatnagar, J.~A. Kennewell, W.~Marquette, J.~Patr{\'o}n, O.~Sa{\'a}, and
  E.~Yasukawa.
\newblock The global oscillation network group (gong) project.
\newblock \emph{Science}, 272\penalty0 (5266):\penalty0 1284--1286, 1996.
\newblock ISSN 0036-8075.
\newblock \doi{10.1126/science.272.5266.1284}.
\newblock URL \url{https://science.sciencemag.org/content/272/5266/1284}.

\bibitem[{Hathaway}(2015)]{Hathaway2015}
David~H. {Hathaway}.
\newblock {The Solar Cycle}.
\newblock \emph{Living Reviews in Solar Physics}, 12\penalty0 (1):\penalty0 4,
  September 2015.
\newblock \doi{10.1007/lrsp-2015-4}.

\bibitem[{Lyot}(1944)]{Lyot1944}
Bernard {Lyot}.
\newblock {Le filtre monochromatique polarisant et ses applications en physique
  solaire}.
\newblock \emph{Annales d'Astrophysique}, 7:\penalty0 31, January 1944.

\bibitem[{Mackay} and {Yeates}(2012)]{Mackay2012}
Duncan~H. {Mackay} and Anthony~R. {Yeates}.
\newblock {The Sun's Global Photospheric and Coronal Magnetic Fields:
  Observations and Models}.
\newblock \emph{Living Reviews in Solar Physics}, 9\penalty0 (1):\penalty0 6,
  November 2012.
\newblock \doi{10.12942/lrsp-2012-6}.

\bibitem[{Mart{\'\i}nez Pillet} et~al.(2011){Mart{\'\i}nez Pillet}, {Del Toro
  Iniesta}, {{\'A}lvarez-Herrero}, {Domingo}, {Bonet}, {Gonz{\'a}lez
  Fern{\'a}ndez}, {L{\'o}pez Jim{\'e}nez}, {Pastor}, {Gasent Blesa}, {Mellado},
  {Piqueras}, {Aparicio}, {Balaguer}, {Ballesteros}, {Belenguer}, {Bellot
  Rubio}, {Berkefeld}, {Collados}, {Deutsch}, {Feller}, {Girela}, {Grauf},
  {Heredero}, {Herranz}, {Jer{\'o}nimo}, {Laguna}, {Meller}, {Men{\'e}ndez},
  {Morales}, {Orozco Su{\'a}rez}, {Ramos}, {Reina}, {Ramos}, {Rodr{\'\i}guez},
  {S{\'a}nchez}, {Uribe-Patarroyo}, {Barthol}, {Gandorfer}, {Knoelker},
  {Schmidt}, {Solanki}, and {Vargas Dom{\'\i}nguez}]{vmp2011}
V.~{Mart{\'\i}nez Pillet}, J.~C. {Del Toro Iniesta}, A.~{{\'A}lvarez-Herrero},
  V.~{Domingo}, J.~A. {Bonet}, L.~{Gonz{\'a}lez Fern{\'a}ndez}, A.~{L{\'o}pez
  Jim{\'e}nez}, C.~{Pastor}, J.~L. {Gasent Blesa}, P.~{Mellado}, J.~{Piqueras},
  B.~{Aparicio}, M.~{Balaguer}, E.~{Ballesteros}, T.~{Belenguer}, L.~R. {Bellot
  Rubio}, T.~{Berkefeld}, M.~{Collados}, W.~{Deutsch}, A.~{Feller},
  F.~{Girela}, B.~{Grauf}, R.~L. {Heredero}, M.~{Herranz}, J.~M.
  {Jer{\'o}nimo}, H.~{Laguna}, R.~{Meller}, M.~{Men{\'e}ndez}, R.~{Morales},
  D.~{Orozco Su{\'a}rez}, G.~{Ramos}, M.~{Reina}, J.~L. {Ramos},
  P.~{Rodr{\'\i}guez}, A.~{S{\'a}nchez}, N.~{Uribe-Patarroyo}, P.~{Barthol},
  A.~{Gandorfer}, M.~{Knoelker}, W.~{Schmidt}, S.~K. {Solanki}, and S.~{Vargas
  Dom{\'\i}nguez}.
\newblock The imaging magnetograph experiment (imax) for the sunrise
  balloon-borne solar observatory.
\newblock \emph{Solar Physics}, 268\penalty0 (1):\penalty0 57--102, January
  2011.
\newblock \doi{10.1007/s11207-010-9644-y}.

\bibitem[Pevtsov et~al.(2020)Pevtsov, Petrie, MacNeice, and
  Virtanen]{Pevtsov2020}
A.~A. Pevtsov, G.~Petrie, P.~MacNeice, and I.~I. Virtanen.
\newblock Effect of additional magnetograph observations from different
  lagrangian points in sun-earth system on predicted properties of quasi-steady
  solar wind at 1 au.
\newblock \emph{Space Weather}, 18\penalty0 (7):\penalty0 e2020SW002448, 2020.
\newblock \doi{https://doi.org/10.1029/2020SW002448}.
\newblock URL
  \url{https://agupubs.onlinelibrary.wiley.com/doi/abs/10.1029/2020SW002448}.
\newblock e2020SW002448 10.1029/2020SW002448.

\bibitem[Sheeley(2005)]{Sheeley2005}
Neil~R. Sheeley.
\newblock Surface evolution of the sun's magnetic field: A historical review of
  the flux-transport mechanism.
\newblock \emph{Living Reviews in Solar Physics}, 2\penalty0 (1):\penalty0 5,
  Oct 2005.
\newblock ISSN 1614-4961.
\newblock \doi{10.12942/lrsp-2005-5}.
\newblock URL \url{https://doi.org/10.12942/lrsp-2005-5}.

\bibitem[{Shepherd} et~al.(1993){Shepherd}, {Thuillier}, {Gault}, {Solheim},
  {Hersom}, {Alunni}, {Brun}, {Brune}, {Charlot}, {Cogger}, {Desaulniers},
  {Evans}, {Gattinger}, {Girod}, {Harvie}, {Hum}, {Kendall}, {Llewellyn},
  {Lowe}, {Ohrt}, {Pasternak}, {Peillet}, {Powell}, {Rochon}, {Ward}, {Wiens},
  and {Wimperis}]{Shepherd1993}
G.~G. {Shepherd}, G.~{Thuillier}, W.~A. {Gault}, B.~H. {Solheim}, C.~{Hersom},
  J.~M. {Alunni}, J.~F. {Brun}, S.~{Brune}, P.~{Charlot}, L.~L. {Cogger}, D.~L.
  {Desaulniers}, W.~F.~J. {Evans}, R.~L. {Gattinger}, F.~{Girod}, D.~{Harvie},
  R.~H. {Hum}, D.~J.~W. {Kendall}, E.~J. {Llewellyn}, R.~P. {Lowe}, J.~{Ohrt},
  F.~{Pasternak}, O.~{Peillet}, I.~{Powell}, Y.~{Rochon}, W.~E. {Ward}, R.~H.
  {Wiens}, and J.~{Wimperis}.
\newblock {WINDII, the WIND imaging interferometer on the Upper Atmosphere
  Research Satellite}.
\newblock \emph{\jgr}, 98\penalty0 (D6):\penalty0 10725--10750, June 1993.
\newblock \doi{10.1029/93JD00227}.

\bibitem[{Shibata} and {Magara}(2011)]{Shibata2011}
Kazunari {Shibata} and Tetsuya {Magara}.
\newblock {Solar Flares: Magnetohydrodynamic Processes}.
\newblock \emph{Living Reviews in Solar Physics}, 8\penalty0 (1):\penalty0 6,
  December 2011.
\newblock \doi{10.12942/lrsp-2011-6}.

\bibitem[{Solanki} et~al.(2020){Solanki}, {del Toro Iniesta}, {Woch},
  {Gandorfer}, {Hirzberger}, {Alvarez-Herrero}, {Appourchaux}, {Mart{\'\i}nez
  Pillet}, {P{\'e}rez-Grande}, {Sanchis Kilders}, {Schmidt}, {G{\'o}mez Cama},
  {Michalik}, {Deutsch}, {Fernandez-Rico}, {Grauf}, {Gizon}, {Heerlein},
  {Kolleck}, {Lagg}, {Meller}, {M{\"u}ller}, {Sch{\"u}hle}, {Staub}, {Albert},
  {Alvarez Copano}, {Beckmann}, {Bischoff}, {Busse}, {Enge}, {Frahm},
  {Germerott}, {Guerrero}, {L{\"o}ptien}, {Meierdierks}, {Oberdorfer},
  {Papagiannaki}, {Ramanath}, {Schou}, {Werner}, {Yang}, {Zerr}, {Bergmann},
  {Bochmann}, {Heinrichs}, {Meyer}, {Monecke}, {M{\"u}ller}, {Sperling},
  {{\'A}lvarez Garc{\'\i}a}, {Aparicio}, {Balaguer Jim{\'e}nez}, {Bellot
  Rubio}, {Cobos Carracosa}, {Girela}, {Hern{\'a}ndez Exp{\'o}sito}, {Herranz},
  {Labrousse}, {L{\'o}pez Jim{\'e}nez}, {Orozco Su{\'a}rez}, {Ramos},
  {Barandiar{\'a}n}, {Bastide}, {Campuzano}, {Cebollero}, {D{\'a}vila},
  {Fern{\'a}ndez-Medina}, {Garc{\'\i}a Parejo}, {Garranzo-Garc{\'\i}a},
  {Laguna}, {Mart{\'\i}n}, {Navarro}, {N{\'u}{\~n}ez Peral}, {Royo},
  {S{\'a}nchez}, {Silva-L{\'o}pez}, {Vera}, {Villanueva}, {Fourmond}, {de
  Galarreta}, {Bouzit}, {Hervier}, {Le Clec'h}, {Szwec}, {Chaigneau},
  {Buttice}, {Dominguez-Tagle}, {Philippon}, {Boumier}, {Le Cocguen},
  {Baranjuk}, {Bell}, {Berkefeld}, {Baumgartner}, {Heidecke}, {Maue}, {Nakai},
  {Scheiffelen}, {Sigwarth}, {Soltau}, {Volkmer}, {Blanco Rodr{\'\i}guez},
  {Domingo}, {Ferreres Sabater}, {Gasent Blesa}, {Rodr{\'\i}guez
  Mart{\'\i}nez}, {Osorno Caudel}, {Bosch}, {Casas}, {Carmona}, {Herms},
  {Roma}, {Alonso}, {G{\'o}mez-Sanjuan}, {Piqueras}, {Torralbo}, {Fiethe},
  {Guan}, {Lange}, {Michel}, {Bonet}, {Fahmy}, {M{\"u}ller}, and
  {Zouganelis}]{Solanki2020}
S.~K. {Solanki}, J.~C. {del Toro Iniesta}, J.~{Woch}, A.~{Gandorfer},
  J.~{Hirzberger}, A.~{Alvarez-Herrero}, T.~{Appourchaux}, V.~{Mart{\'\i}nez
  Pillet}, I.~{P{\'e}rez-Grande}, E.~{Sanchis Kilders}, W.~{Schmidt}, J.~M.
  {G{\'o}mez Cama}, H.~{Michalik}, W.~{Deutsch}, G.~{Fernandez-Rico},
  B.~{Grauf}, L.~{Gizon}, K.~{Heerlein}, M.~{Kolleck}, A.~{Lagg}, R.~{Meller},
  R.~{M{\"u}ller}, U.~{Sch{\"u}hle}, J.~{Staub}, K.~{Albert}, M.~{Alvarez
  Copano}, U.~{Beckmann}, J.~{Bischoff}, D.~{Busse}, R.~{Enge}, S.~{Frahm},
  D.~{Germerott}, L.~{Guerrero}, B.~{L{\"o}ptien}, T.~{Meierdierks},
  D.~{Oberdorfer}, I.~{Papagiannaki}, S.~{Ramanath}, J.~{Schou}, S.~{Werner},
  D.~{Yang}, A.~{Zerr}, M.~{Bergmann}, J.~{Bochmann}, J.~{Heinrichs},
  S.~{Meyer}, M.~{Monecke}, M.~F. {M{\"u}ller}, M.~{Sperling}, D.~{{\'A}lvarez
  Garc{\'\i}a}, B.~{Aparicio}, M.~{Balaguer Jim{\'e}nez}, L.~R. {Bellot Rubio},
  J.~P. {Cobos Carracosa}, F.~{Girela}, D.~{Hern{\'a}ndez Exp{\'o}sito},
  M.~{Herranz}, P.~{Labrousse}, A.~{L{\'o}pez Jim{\'e}nez}, D.~{Orozco
  Su{\'a}rez}, J.~L. {Ramos}, J.~{Barandiar{\'a}n}, L.~{Bastide},
  C.~{Campuzano}, M.~{Cebollero}, B.~{D{\'a}vila}, A.~{Fern{\'a}ndez-Medina},
  P.~{Garc{\'\i}a Parejo}, D.~{Garranzo-Garc{\'\i}a}, H.~{Laguna}, J.~A.
  {Mart{\'\i}n}, R.~{Navarro}, A.~{N{\'u}{\~n}ez Peral}, M.~{Royo},
  A.~{S{\'a}nchez}, M.~{Silva-L{\'o}pez}, I.~{Vera}, J.~{Villanueva}, J.~J.
  {Fourmond}, C.~Ruiz {de Galarreta}, M.~{Bouzit}, V.~{Hervier}, J.~C. {Le
  Clec'h}, N.~{Szwec}, M.~{Chaigneau}, V.~{Buttice}, C.~{Dominguez-Tagle},
  A.~{Philippon}, P.~{Boumier}, R.~{Le Cocguen}, G.~{Baranjuk}, A.~{Bell}, Th.
  {Berkefeld}, J.~{Baumgartner}, F.~{Heidecke}, T.~{Maue}, E.~{Nakai},
  T.~{Scheiffelen}, M.~{Sigwarth}, D.~{Soltau}, R.~{Volkmer}, J.~{Blanco
  Rodr{\'\i}guez}, V.~{Domingo}, A.~{Ferreres Sabater}, J.~L. {Gasent Blesa},
  P.~{Rodr{\'\i}guez Mart{\'\i}nez}, D.~{Osorno Caudel}, J.~{Bosch},
  A.~{Casas}, M.~{Carmona}, A.~{Herms}, D.~{Roma}, G.~{Alonso},
  A.~{G{\'o}mez-Sanjuan}, J.~{Piqueras}, I.~{Torralbo}, B.~{Fiethe}, Y.~{Guan},
  T.~{Lange}, H.~{Michel}, J.~A. {Bonet}, S.~{Fahmy}, D.~{M{\"u}ller}, and
  I.~{Zouganelis}.
\newblock {The Polarimetric and Helioseismic Imager on Solar Orbiter}.
\newblock \emph{\aap}, 642:\penalty0 A11, October 2020.
\newblock \doi{10.1051/0004-6361/201935325}.

\bibitem[{Title} and {Ramsey}(1980)]{Title1980}
Alan~M. {Title} and H.~E. {Ramsey}.
\newblock {Improvements in birefringent filters. 6: Analog birefringent
  elements}.
\newblock \emph{\ao}, 19\penalty0 (12):\penalty0 2046--2058, June 1980.
\newblock \doi{10.1364/AO.19.002046}.

\bibitem[{van Driel-Gesztelyi} and {Green}(2015)]{vandriel2015}
Lidia {van Driel-Gesztelyi} and Lucie~May {Green}.
\newblock {Evolution of Active Regions}.
\newblock \emph{Living Reviews in Solar Physics}, 12\penalty0 (1):\penalty0 1,
  September 2015.
\newblock \doi{10.1007/lrsp-2015-1}.

\bibitem[{Wiegelmann} and {Sakurai}(2012)]{Wiegelmann2012}
Thomas {Wiegelmann} and Takashi {Sakurai}.
\newblock {Solar Force-free Magnetic Fields}.
\newblock \emph{Living Reviews in Solar Physics}, 9\penalty0 (1):\penalty0 5,
  September 2012.
\newblock \doi{10.12942/lrsp-2012-5}.

\end{thebibliography}

\end{document}